\documentclass{article}

\usepackage[utf8]{inputenc}
\usepackage[T1]{fontenc}
\usepackage{fouriernc}
\usepackage[french,english]{babel}
\usepackage{amsthm,amssymb,amsfonts,amsmath,algorithm,algorithmic,color,graphicx,rotating,float,url,bbm,multirow,caption,subfig,placeins,xspace,eucal}
\usepackage{amsfonts}
\usepackage{graphicx}
\usepackage{epstopdf}
\usepackage{latexsym,amsmath,color}
\usepackage{booktabs}
\usepackage{bm}
\usepackage{url}
\usepackage{algorithm}

\usepackage[pagebackref = true]{hyperref}
\hypersetup{
  colorlinks = true,
  urlcolor = blue, 
  linkcolor = blue,
  citecolor = blue,
  pdftitle = {No title - \today},
  pdfauthor = {No author},
  pdfsubject = {No subject}
}

\def\1{\mathbbm{1}}

\addto\extrasenglish{%
    
}
\def\Rit{\mathbb{R}}

\def\1{\mathbb{I}}

\def\Prob{\mathbb{P}}

\graphicspath{{../fig/}}

\makeatletter
\newcommand{\vast}{\bBigg@{4}}
\newcommand{\Vast}{\bBigg@{5}}
\makeatother

\title{Improving prediction performance of stellar\\ parameters using functional models}
\author{Sylvain Robbiano\footnote{CIMFAV-Facultad de Ingenier\'ia, Universidad de Valpara\'iso,
Chile, \href{mailto:sylvain.robbiano@gmail.com}{sylvain.robbiano@gmail.com}}, Matthieu Saumard\footnote{Pontificia Universidad Cat\'olica de Valpara\'iso, Instituto de Estad\'istica, 
Chile} and  Michel Cur\'e\footnote{Instituto de F\'isica y Astronom\'ia, Facultad de Ciencias,  Universidad de Valpara\'iso,
Chile}}

\begin{document}
\maketitle

\begin{abstract}
This paper investigates the problem of prediction of stellar parameters, based on the star's electromagnetic spectrum. The knowledge of these parameters permits to infer on the evolutionary state of the star. From a statistical point of view, the spectra of different stars can be represented as functional data. Therefore, a two-step procedure decomposing the spectra in a functional basis combined with a regression method of prediction is proposed. We also use a bootstrap methodology to build prediction intervals for the stellar parameters. A practical application is also provided to illustrate the numerical performance of our approach.
\end{abstract}

\section{Introduction}

%
In stellar astrophysics, one of the major issues is to obtain information about stars from the observation of their spectra. A spectrum can be mathematically considered as a functional object where the flux (the physical observable quantity)  is expressed as function of the wavelength or frequency (see e.g., \cite{Rob07}). From the spectrum, an astrophysicist can deduce some  of the stellar parameters such as the radius, temperature, gravity of the star, as well as the abundances of some of its elements. A standard method (see e.g., \cite{KP2000,Lefever07}) consists of modelling (using state-of-the-art numerical codes) many theoretical spectra for each individual star and carefully comparing these by eye with the star's observed spectrum, in order to get the stellar parameters. \\
 Due to the growing number of new astronomical facilities and instrumentation, it is now possible to obtain the observed spectra for hundreds of stars simultaneously (e.g., using the Flames, Uves or Fors multi-spectral object instruments  in ESO-VLT at Cerro Paranal in northen Chile).  Thus, when using all these data to obtain the stellar parameters, it is very important to use novel techniques that significantly reduce the data processing time.  \\
In this context, different approaches have been developed in recent years.  
On one side  \cite{Mokiem05} implemented a genetic algorithm \cite{pikaia} to develop an automated fitting method for the quantitative spectroscopy of O- and early B-type stars with stellar winds. On the other side \cite{Lefever07,IACOB}  created a grid of models for the analysis of massive blue supergiant stars. After the grid is completed, one compares some of the observed spectral lines (see below) with each one of the lines  from all synthetic models in order to get the stellar parameters.\\ 
Using a grid of models we want to speed up the process of obtaining the closed synthetic spectra using functional data analysis.\\
Due to the functional form of this type of data, we choose to create prediction methods using the tools developed by the functional data analysis community (see e.g., the monographs of \cite{ramsay} and \cite{ferraty}).\\  This adds a new field of application for functional data analysis to the existing ones on others areas, e.g., spectrometry, neuroscience, econometrics. Functional data have been used in several contexts such as classification (see \cite{ferraty}, \cite{cuevas}, \cite{cuevas2}, \cite{del1}, \cite{del2}) and prediction (see \cite{del3}, \cite{hsing}). To the best of our knowledge, there is only one article that adapts the functional data methodology to astronomy, see \cite{ciol}.\\
In this article, we tackle two problems that are crucial for astrophysics. The first objective is to predict the star's characteristics using synthetic spectra while the second objective is to create prediction intervals for the obtained stellar parameters. In a first step, we decompose the spectrum into a functional basis that can be Fourier or splines. Then we use a prediction method to determine the values of the stellar parameters, specifically linear regression and robust or penalized linear regression. In order to choose the number of bases in the decomposition, we propose to compute the accuracy of our predictions using a cross-validation method. We also develop a method based on the bootstrap principle in order to obtain prediction intervals for the stellar parameters. These two automatic procedures can reduce significantly the number of simulated spectra in order to obtain information about these parameters.\\
This paper is organized as follows: In section \ref{data}, we present the dataset used throughout the paper as well as the classical method used by astrophysicists to predict stellar parameters. In section \ref{Funct}, we recall tools developed for functional data analysis and we present methodology for decomposing the spectrum. In section \ref{Eval}, we present the prediction methods and we propose two procedures to evaluate the quality of the prediction and the computation of the prediction intervals. Section \ref{numeric} is dedicated to explanation of the numerical results and last section discuss our conclusions.

\section{Data}\label{data}
\begin{figure}[!t]

\centering
\begin{center}
\includegraphics[width=12cm]{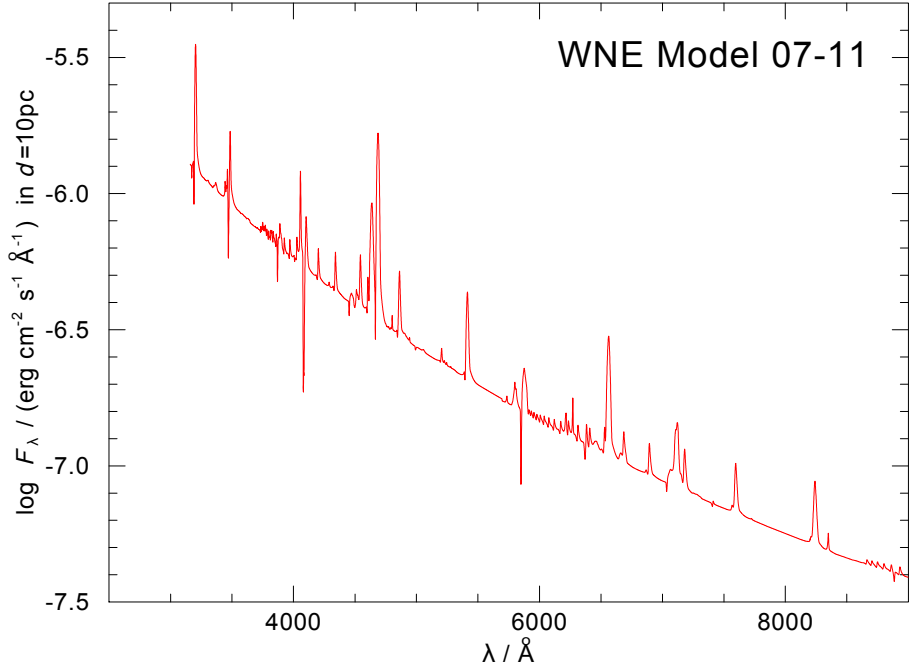}
\caption{\label{curve} Synthetic spectra of a WN star (grid model 07-11). Physical flux in cgs 
units is plotted versus wavelength in Angstrom. The spectra is the one a star would have if is located at a distance of 10 parsecs (pc) from us.}
\end{center}
\end{figure}

\begin{figure}[!t]
\centering
\begin{center}
\includegraphics[width=12cm]{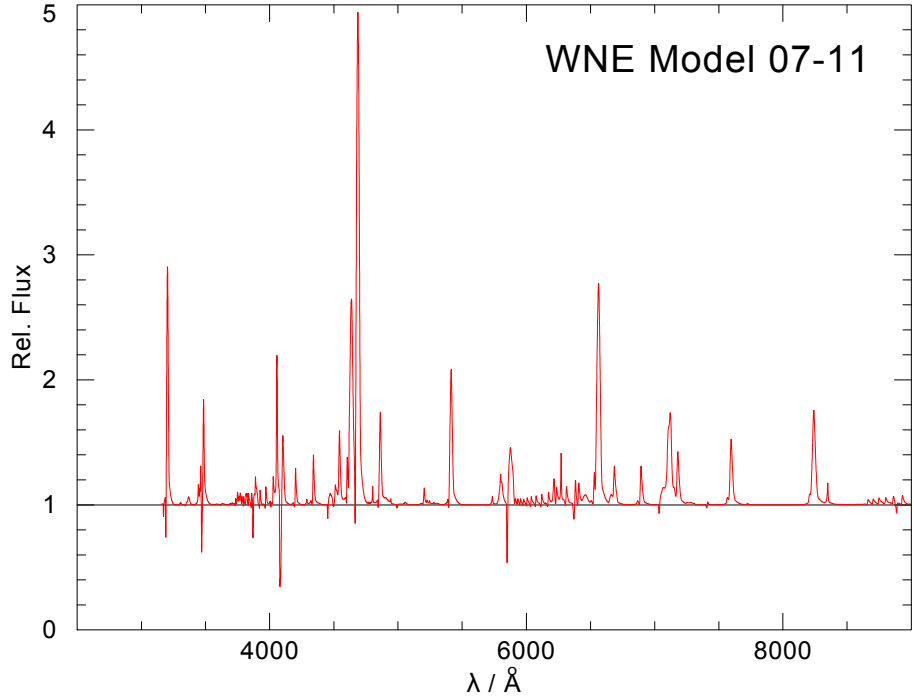}
\caption{\label{curv-norm} Same as Fig. 1 but the flux is normalized by the flux of the continuum, 
see text for details.}
\end{center}

\end{figure}

\begin{figure}[!t]

\centering
\begin{center}
\includegraphics[width=12cm]{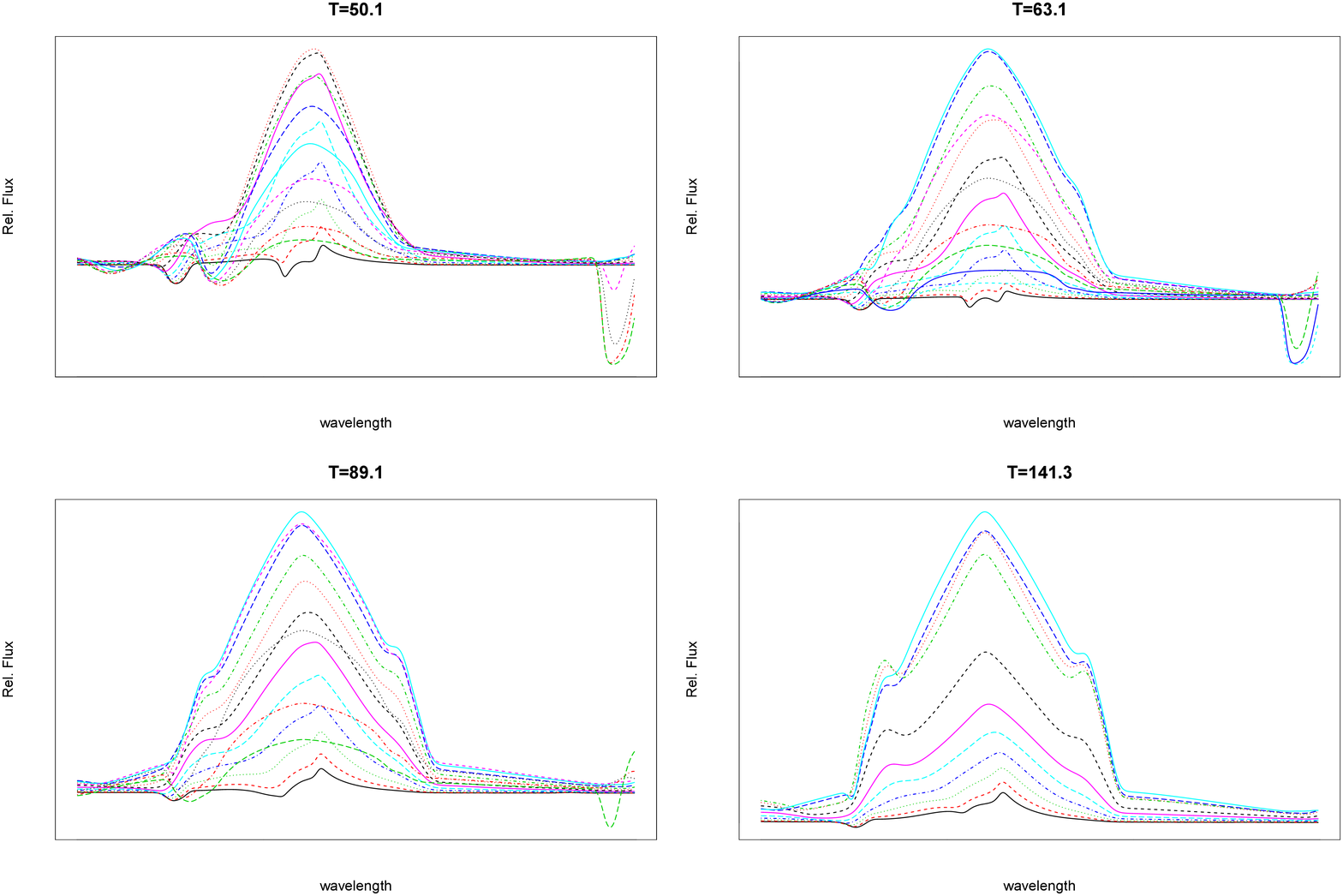}
\caption{\label{fig:Yfix} $H-\alpha$ line (6562.8 angstrom) plotted against wavelength in the range $[6500, 6600]$. Upper left panel plot this line for models at Temperature=50.1 kKelvin showing the dependence for different stellar radius $\log (R_t)$ in the range $[0.3,1.7]$. Other panels show the same but for different effective temperatures.}
\end{center}
\end{figure}

\begin{figure}[!t]
\centering
\begin{center}
\includegraphics[width=12cm]{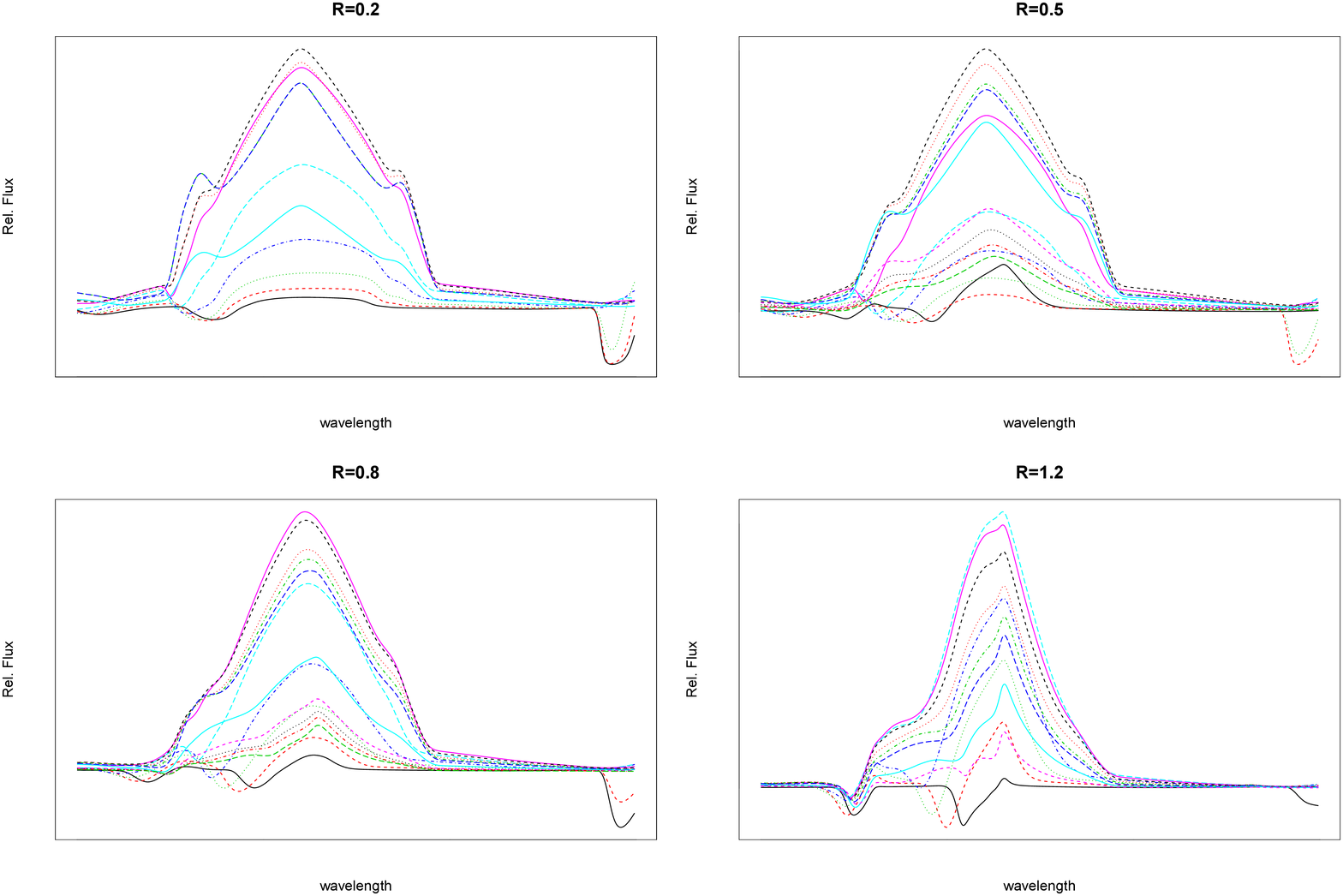}
\caption{\label{fig:Zfix} Same as Fig 3. Upper left panel plot this line for models at $\log(R_t)=0.2$, but here the dependence is function of the effective temperature in the range $[56.2,177.8]$. Other panels show the same but for different $\log(R_t)$.}
\end{center}

\end{figure}

We use a grid of models calculated using the numerical code PoWR, \cite{hamann04}. This code calculates the synthetic spectrum by solving the radiative transport equation for a stellar atmosphere. From a physical point of view, the observed spectra measures the flux in the outer layers of a star (photosphere) and has two components (for details, see \cite{Rob07}): the continuum (corresponding to the deepest layers of the photosphere) and the spectral lines (interaction of the radiation with non full--ionized atoms). We are interested in comparing some observed spectral lines with the corresponding lines in synthetic spectra to get the information about the stellar parameters. Thus, we divide the observed spectrum by the continuum to get what astronomers call the 'normalized spectrum', which only carries information about the spectral lines.
From the PoWR database we will therefore use the normalized line spectrum of a star. All data can be downloaded at the webpage \url{http://www.astro.physik.uni-potsdam.de/~wrh/ PoWR/powrgrid2-WNE.html}.  There are many variables that describe the evolution of a star; in this case we will use just 2 of them, namely the modified radius, $R_{t}$, and the effective temperature,  $T_{*}$, (see details in \cite{hamann04}). We downloaded the normalized spectra, each of which contains 10455 data points over the same grid of wavelengths. The spectra were generated for Wolf-Rayet star of type N (WNE). \\
The WNE database contains modelled spectra for 210 parameter pairings, corresponding to $T_*\in [31.6,199.5]$ and $R_t\in [0,1.5]$. As the data are on a grid, several spectra share the same value of $T_{*}$ or $R_{t}$.  We compare ten spectral lines that correspond to Hydrogen and Helium, the most abundant elements. Specifically the lines are at wavelengths in the intervals $\left[6500,6600\right],\,\left[4800,4900\right],\,\left[4300,4360\right],$ $\,\left[4370,4405\right],\,\left[4460,4485\right],\, \left[4700,4730\right], 
\left[4900, 4940\right],  \left[4180,4220\right],\,\left[4525,4560\right],$ $\,\left[4670,4700\right].$ In Figures \ref{curve} and \ref{curv-norm}, we plot the whole spectrum corresponding to point 07-11 in the grid of models. In Figures \ref{fig:Yfix} and \ref{fig:Zfix}, we plot the modelled line spectra in the interval $\left[6500,6600\right]$ with one of the parameters fixed, to show the difficulty of the problem. This grid is used to predict the value of $T_*$ and $R_t$ when observing a star through its spectrum. The usual approach to prediction involves minimizing the $L_2$ distance between the observed line spectrum and the line spectra generated from the grid of parameters, and choosing the parameters of the minimizer.

\section{Functional linear model}\label{Funct}
Due to high correlations between two consecutive values of a spectrum, traditional tools for parameter prediction, such as multivariate linear regression, fail. To overcome this problem, we take advantage of the functional nature of the objects. Functional data analysis has become an important field of modern statistics, and now there exists an abundant literature on this topic. We choose in particular the approach of the functional linear model. We have tried nonparametric methods such as in \cite{ferraty} in a preliminary study, but it seems that the functional linear model works better.
Functional linear regression has been introduced for the case where predictors are curves, which is exactly the case here. In fact, the functional linear model can be seen as a continuous version of the multivariate linear model when the predictor is functional, i.e., where the predictor takes values in a functional space. The  functional linear model, due to its simplicity, has been widely studied in the literature; we refer to \cite{cardot99} for a description of the model and an approach to parameter estimation. Furthermore, the functional linear model covers a large range of applications, see \cite{ramsay} for examples.  Our case corresponds to a functional multivariate setup and we recall here the principle and the notation, see \cite{matsui08} for more details.\\
Suppose we have $n$ observations i.i.d. $\left(Y^{(1)}_{i},Y^{(2)}_i, X_{i}\right)$; $i=1,\ldots,n$, of \\$(Y^{(1)},Y^{(2)},X)$ where $Y^{(1)}$ and $Y^{(2)}$ are scalar responses and $X$ is a functional predictor. 
Suppose also that $X$ is observed at $M$ locations of interest, such that we have $X_{i,j}$, $j=1,\ldots,M$. In our problem, $Y^{(1)}$ and $Y^{(2)}$ represent the stellar parameters $T_*$ and $R_t$, respectively, and the functional variable $X$ represents the spectrum of the considered star, which we observe at $M=10$ spectral lines.
The model is a multivariate functional linear model with scalar responses and is given by:
\begin{equation}\label{model1}
\left\{
 \begin{array}{ll}
	Y^{(1)}_i = \alpha_1+ \sum_{j=1}^M\langle \beta_{1,j}, X_{i,j}\rangle +\varepsilon_i \\
        Y^{(2)}_i =\alpha_2+\sum_{j=1}^M \langle \beta_{2,j}, X_{i,j} \rangle +\varepsilon'_i
 \end{array}
\right.
\end{equation}
where $\alpha_1$ and $\alpha_2$ are unknown scalar parameters,  $\beta_{1,j}$ and $\beta_{2,j}$ are unknown functional parameters, and $\varepsilon_i$ and $\varepsilon'_i$ are i.i.d. errors independent of the predictors. The notation $ \langle \cdot , \cdot \rangle $ represents the inner product of the space $L^2(K_j)$, which is the space of square integrable functions on compact intervals $K_j$, which represent the $M=10$ different intervals (containing the 10 spectral lines, see section \ref{data}).\\
In order to estimate the functional parameters, we use a Fourier basis. We denote the $p$-length vector of Fourier basis functions $(\rho^j_1 (t),\ldots, \rho^j_p (t))$. The functional predictor $X_{i,j}$ then has the following basis function representation:
\begin{equation*}
X_{i,j} = \sum_{k=1}^p x_{ijk} \rho^j_k+r_{ijp}, 
\end{equation*} 
where $x_{ijk}=\int_K X_{i,j} (t) \rho^j_k (t) dt$ and $r_{ijp}$ is an error. We decompose the parameters $\beta_{1,j}$ and $\beta_{2,j}$ using the same basis with the same notation. Then, we have
\begin{equation}\label{model2}
\left\{
 \begin{array}{ll}
	Y^{(1)}_i = \alpha_1+\sum_{j=1}^M \sum_{k=1}^p \beta_{1jk} x_{ijk}  +\epsilon_i \\
        Y^{(2)}_i =\alpha_2+\sum_{j=1}^M \sum_{k=1}^p \beta_{2jk}x_{ijk}  +\epsilon'_i.
 \end{array}
\right.
\end{equation}
The errors in the equations of model (\ref{model1}) are different to the errors in (\ref{model2}), due to the decomposition of the predictors. 
If we denote by $\bm{X_j}$ the matrix $(x_{ijk})_{i=1\ldots n,k=1,\ldots,p}$ we have
\begin{equation}
\left\{
 \begin{array}{ll}
	\bm{Y^{(1)}} = \alpha_1+\sum_{j=1}^M \bm{X_j}\beta_{1,j}  +\bm{\epsilon} \\
        \bm{Y^{(2)}} = \alpha_2+\sum_{j=1}^M \bm{X_j}\beta_{2,j}    +\bm{\epsilon}'.
 \end{array}
\right.
\end{equation}
\\
Given the new variable $x$, in order to predict values for $\widehat{Y}^{(1)}$ and $\widehat{Y}^{(2)}$, we need to estimate the unknown parameters. Prediction is then simply achieved through plugging-in these estimates:
\begin{equation*}
\left\{
 \begin{array}{ll}
	\widehat{Y}^{(1)} = \widehat{\alpha}_1+ \sum_{j=1}^M\langle \widehat{\beta}_{1,j}, x_j \rangle \\
        \widehat{Y}^{(2)}=\widehat{\alpha}_2+ \sum_{j=1}^M\langle \widehat{\beta}_{2,j}, x_j \rangle 
 \end{array}
\right.
\end{equation*}

In the literature, there are few references about the theoretical selection of the numbers of components $p$, see \cite{Selec-hall} and \cite{selec-LWC}. This is a challenging issue and in our approach, we choose to minimize the $\Gamma$ME, see section \ref{numeric}.

\section{Evaluation methods}\label{Eval}
In this section, we present the statistical methods that we use to predict the stellar parameters, given the data matrix ${\bm X}$ obtained through the functional models.

\subsection{Notations}
We denote the $l_{2}$ norm by $\|.\|_{2}$, i.e. $\|\beta\|^2_{2}=\sum^{p}_{k=1}|\beta_{k}|^{2}$. Similarly, we denote the $l_{1}$ norm by $\|.\|_{1}$, i.e. $\|\beta\|_{1}=\sum^{p}_{k=1}|\beta_{k}|$.
In order to measure the quality of the prediction function, we use the root mean square error i.e 
$$ RMSE(Y^{(l)},\widehat{Y}^{(l)})=\sqrt{\frac{1}{n}\sum^{n}_{i=1}|Y_{i}^{(l)}-\widehat{Y}_{i}^{(l)}|^{2}}.
$$
Of course, the goal is to obtained a RMSE as small as possible.
We also introduce the covariance matrix $\Gamma=(\gamma_{l,p})_{1\leq l,p\leq 2}$ where $\gamma_{l,p} = cov(Y^{(l)},Y^{(p)})$ and the associated scalar product $(u,v)_{\Gamma}=u^{T}\Gamma^{-1}v$. Let us define the  mean $\Gamma$ error, i.e
$$ \Gamma ME(Y,\widehat{Y})=\sqrt{\frac{1}{n}\sum^{n}_{i=1}(Y_{i}-\widehat{{Y}_{i}},Y_{i}-\widehat{{Y}_{i}})_{\Gamma}}.
$$

\subsection{Linear regression and variations}

The easiest approach to prediction is to fit a linear model for the parameters. This involves finding the parameter vector $\beta$ that minimizes $\|\bm{X}\beta-\bm{Y}^{(l)}\|_2 $ and classical results show that the required minimizer is $\hat{\beta}^{OLS}=(\bm{X}^{T}\bm{X})^{-1}\bm{X}\bm{Y}^{(l)}$. This method suffers from a major drawback, however: it is very unstable when the matrix ${\bf X^T X}$ is ill-conditioned. Moreover, when the number of basis functions becomes large, the matrix ${\bf X^T X}$ is no longer invertible; in our setting we have 126 curves so this is the case when $p>13$. In order to obtain more consistent predictions, we explore three variants of this method. The first one is called ``robust linear regression", see e.g., \cite{huber1964}. This method consists of solving an M-estimation problem in order to stabilize the linear regression when the noise is heavy-tailed, but this method is not suited for large $p$. The second solution is to penalize the linear regression: specifically, we minimize the equation  $\|\bm{Y}^{(l)}-\bm{X}\beta\|_{2} +\lambda \|\beta\|^2_{2}$ and this method is called ridge regression \cite{Hoerl1}. One can show that the coefficient can be written as $\hat{\beta}_{j}^{RR}=\frac{\hat{\beta}_{j}^{OLS}}{1+\lambda}$.  When the number of parameters is larger than the number of observations, the classical approach is to use a lasso procedure, see \cite{tibshirani96regression}. This is the third method that we consider, and it consists of minimizing $\|\bm{Y}^{(l)}-\bm{X}\beta\|_{2} +\lambda \|\beta\|_{1}$.

\subsection{Evaluation of the prediction}\label{subsec:4.2}
Here, we describe how to evaluate the accuracy of the presented algorithms through a classical cross validation procedure. We choose to use a 5-fold procedure; 3 folds are used to estimate the parameter vector $\beta^{(l)}$. Notice that the functional model depends on the number, $p$, of basis functions used for the decomposition. To choose this parameter, we use a validation set as described in Algorithm \ref{algo:VC}. Since we want to use the same parameter $p$ to estimate $\hat{\beta_{1}}$ and $\hat{\beta_{2}}$, we choose the parameter $p$ that minimizes the mean $\Gamma$ error on the validation set. Finally, the performance of an algorithm is evaluated on the test set. In order to get a more stable evaluation of the performances, we circulate the folds and average the results. We notice that the error for each parameter is sensitive with respect to the number of basis functions used, and for this reason we also try to standardize the vector of stellar parameters within each column of the training set.

\begin{algorithm}[ht!]
\begin{center}
\caption{\label{algo:VC}  Evaluation of the prediction with validated number of basis}
{\small

\begin{enumerate}
\item ({\sc Input.}) A dataset $\mathcal{D}$, $\mathcal{A}$ a prediction algorithm.
\medskip

\item ({\sc Split}) Split $\mathcal{D}_n$ randomly in five equal parts $\mathcal{D}_1$, $\mathcal{D}_2$, $\mathcal{D}_3$, $\mathcal{D}_4$ and $\mathcal{D}_5$. Define $\mathcal{D}_a=\cup^{3}_{i=1}\mathcal{D}_i $, $\mathcal{D}_v=\mathcal{D}_4$, $\mathcal{D}_t=\mathcal{D}_5$
\medskip

\item ({\sc Iterations.}) For $p=1,3,5\; \ldots,13$, 
\begin{enumerate}
\item ({\sc Decomposition.}) Decompose $\mathcal{D}_a$, $\mathcal{D}_v$ using a functional basis, $\mathcal{F}$, of size $p$ to obtain the matrices $\textbf{X}_a$ and $\textbf{X}_v$.
\item ({\sc Learning.}) Use $\mathcal{A}$ and $\textbf{X}_a$ to compute $\widehat{\beta}^{p,1}$ and $\widehat{\beta}^{p,2}$.
\item ({\sc Validation.}) Compute $\widehat{\bm{Y}}_v^{p,1}=\textbf{X}_v\widehat{\beta}^{p,1}$ and $\widehat{\bm{Y}}_v^{p,2}=\textbf{X}_v\widehat{\beta}^{p,2}$ and $\Gamma ME (\bm{Y}_{v},\widehat{\bm{Y}}^{p}_{v})$
where $\bm{Y}_{v}$ is the matrix of true parameters corresponding to the validation set and $\widehat{\bm{Y}}_v^{p}=(\widehat{\bm{Y}}_v^{p,1},\widehat{\bm{Y}}_v^{p,2})$
\end{enumerate}
\medskip

\item ({\sc Prediction}) Choose $\tilde{p}=arg\min_{p}\Gamma ME (\bm{Y}_{v},\widehat{\bm{Y}}^{p}_{v})$. Decompose $\mathcal{D}_a$, $\mathcal{D}_t$ using a functional basis, $\mathcal{F}$, of size $\tilde{p}$ to obtain the matrices $\textbf{X}_a$ and $\textbf{X}_t$. Use $\mathcal{A}$ and $\textbf{X}_a$ to compute $\widehat{\beta}^{\tilde{p},1}$ and $\widehat{\beta}^{\tilde{p},2}$. Compute $\widehat{\bm{Y}}_t^{\tilde{p},1}=\textbf{X}_t\widehat{\beta}^{\tilde{p},1}$ and $\widehat{\bm{Y}}_t^{\tilde{p},2}=\textbf{X}_t\widehat{\beta}^{\tilde{p},2}$. We note $\widehat{\bm{Y}_{t}}=(\widehat{\bm{Y}}_t^{\tilde{p},1},\widehat{\bm{Y}}_t^{\tilde{p},2})$.
\item ({\sc Output.}) $RMSE(\bm{Y}_{t}^{(1)},\widehat{\bm{Y}}_{t}^{\tilde{p},1})$, $RMSE(\bm{Y}_{t}^{(2)},\widehat{\bm{Y}}_{t}^{\tilde{p},2})$, $\Gamma ME(\bm{Y}_{t},\widehat{\bm{Y}_{t}})$ where $\bm{Y}_{t}=(\bm{Y}^{(1)}_{t},\bm{Y}^{(2)}_{t})$ are the true parameters corresponding to the test set.
\end{enumerate}
}
\end{center}
\end{algorithm}

\subsection{Prediction intervals}\label{sec:2.3}
Once we have predicted the parameter values, it is important to say how confident we are about this prediction. To solve this issue we build prediction intervals at level $\alpha\in [0,1]$ for the parameters. By definition, a prediction interval $I$ at level $\alpha$ is such that $\Prob\{Y\in I\}=1-\alpha$, where $Y$ is a random variable. Note that there are a large number of intervals that satisfy this condition, so we focus on prediction interval of the following form: $I=I_{l}\cap I_{r}$ where $I_{l}$ is a left prediction interval  (i.e, of the form $]-\infty; M]$, $M\in \Rit$) at level  $\alpha/2$ and $I_{r}$ is a right  prediction interval  (i.e, of the form $]m; +\infty ]$, $m\in \Rit$) at level  $\alpha/2$. Obviously, such an interval is a prediction interval at level $\alpha$ and is unique. Since the distribution of the parameters $Y^{(k)}$, $k=1,2$, are unknown, the question is how to build a prediction  interval. In this paper, we use the bootstrap methodology (described in \cite{Stine85}) to build a prediction interval for $Y=\theta(X)$ and we recall the procedure in Algorithm \ref{algo:PI}. The idea behind this method is to re-sample from the residuals in order to create the prediction intervals. In order to assess the accuracy of the prediction interval, we estimate the coverage probabilities, i.e., the frequency at which the true parameter falls into the prediction interval. In practice, we run a leave one out procedure to evaluate the coverage probabilities, each a element of the grid is used once as $X_{f}$ and  $\mathcal{D}$ is the rest of the grid in Algorithm \ref{algo:PI}.

\begin{algorithm}[ht!]
\begin{center}
\caption{\label{algo:PI} Bootstrap for building prediction intervals}

{\small

\begin{enumerate}
\item ({\sc Input.}) A training dataset $\mathcal{D}=(\textbf{X},\bm{Y})$ of size $n$, $\mathcal{A}$ a prediction algorithm, $X_{f}$ an observation.
\medskip

\item ({\sc Estimation}) Compute $\widehat{\bm{Y}}=\textbf{X}\widehat{\beta}$ using $\mathcal{D}$ and $\mathcal{A}$. Set $\widehat{r}=(\widehat{r}_{1},\ldots,\widehat{r}_{n})=\bm{Y}-\widehat{\bm{Y}}$ the residuals.
\medskip

\item ({\sc Iterations.}) For $t=1,\; \ldots, T$, 
\begin{enumerate}
\item ({\sc Sample.}) Draw with replacement from $\widehat{r}$ a sample $r^{*}=(r_{1}^{*},\ldots,r^{*}_{n})$ and $r_{f}^{*}$. Create $\mathcal{D}^{*}=(\textbf{X},\bm{Y}^{*})$ where $\bm{Y}^{*}=\textbf{X}\widehat{\beta}+r^{*}$
\item ({\sc Learning.}) Use $\mathcal{A}$ and $\mathcal{D}^{*}$ to compute $\widehat{\beta}^{*}$. Keep $B_{t}^{*}=X_{f}\widehat{\beta}-X_{f}\widehat{\beta}^{*}+r_{f}^{*}$.
\end{enumerate}
\medskip

\item ({\sc Output}) A prediction interval $\tilde{I}=[X_{f}\widehat{\beta}+ q^{*};X_{f}\widehat{\beta}+Q^{*}]$ where $q^{*}$ (resp. $Q^{*}$) is the empirical quantile at level $\alpha/2$ (resp. $1-\alpha/2 $) of $B^{*}=(B_{1}^{*},\dots,B_{T}^{*})$.
\end{enumerate}
}

\end{center}
\end{algorithm}

\section{Numerical results}\label{numeric}

To implement the experiments in this paper we use several R-packages. The functional decomposition is carried out using the package ``fda" , the linear regression (named ``LM" in the Tables), the robust linear regression (named ``Robust LM" in the Tables) and the ridge regression (named ``Ridge" in the Tables) are implemented in the ``MASS" library , and finally the lasso is implemented using the ``glmnet'' library. The standard method used by astrophysicists is named ``Astro" in the tables. As explained in \ref{subsec:4.2}, we run the experiments with the original vectors $Y$ (named ``Brut'' in the Tables)  and with normalised $Y$ (named ``Norm'' in the Tables). Since the matrix $\Gamma$ is unknown we evaluate it using the full database and we obtain $\hat{\Gamma}=(0.2316 ,-0.1119; -0.1119 , 0.2184)$.

\subsection{Selection of the number of basis}

\begin{figure}[t]
\begin{center}

\caption{\label{figSelp} Evolution of the $\Gamma$ME as function of the number of basis for the different 
prediction procedures, see text for details.}
\begin{tabular}{cc}
\parbox{8cm}{
\begin{center}
\includegraphics[width=8cm]{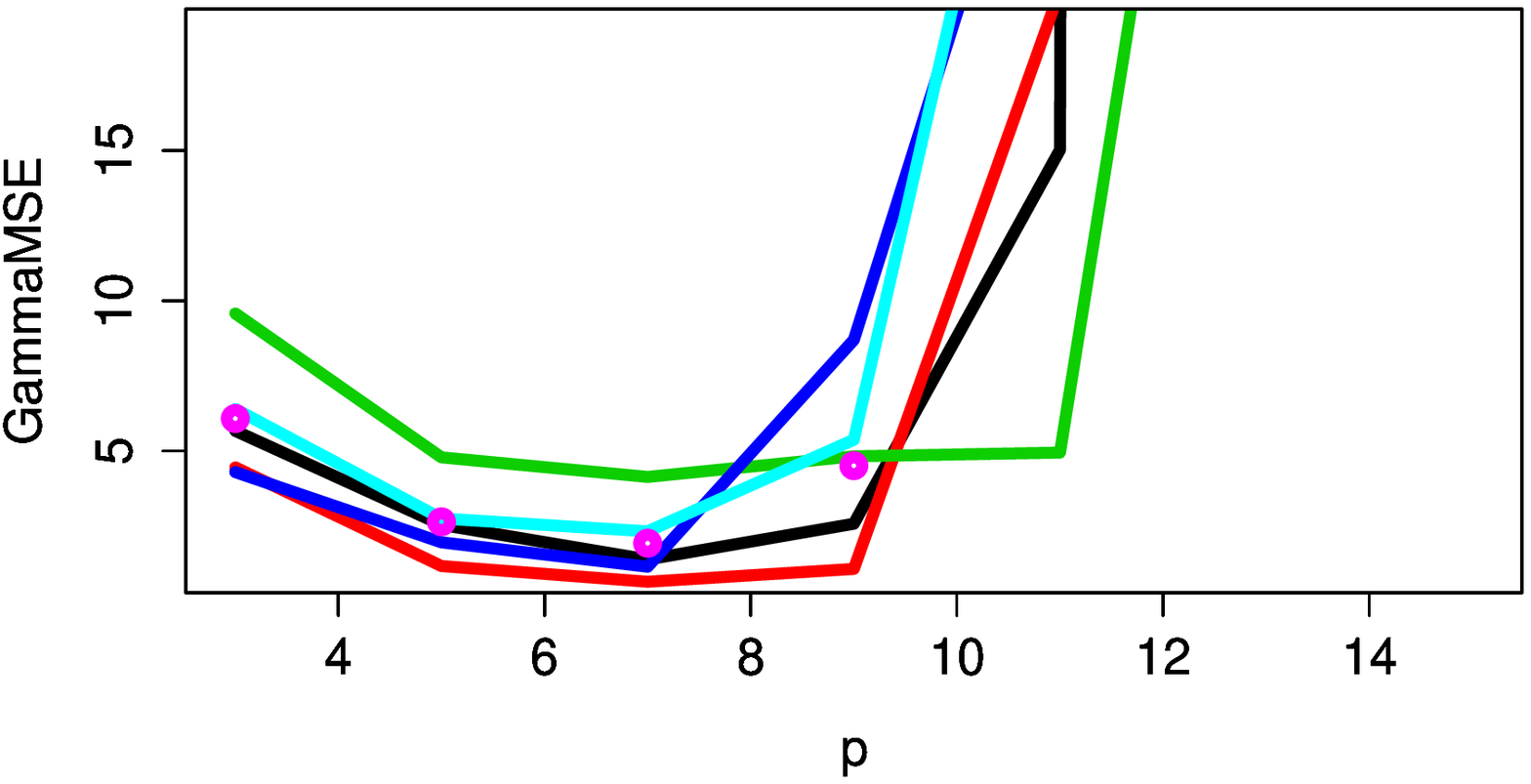}
\tiny{Selection of $p$ for the linear model. }
\end{center}
}&
\parbox{8cm}{
\begin{center}
\includegraphics[width=8cm]{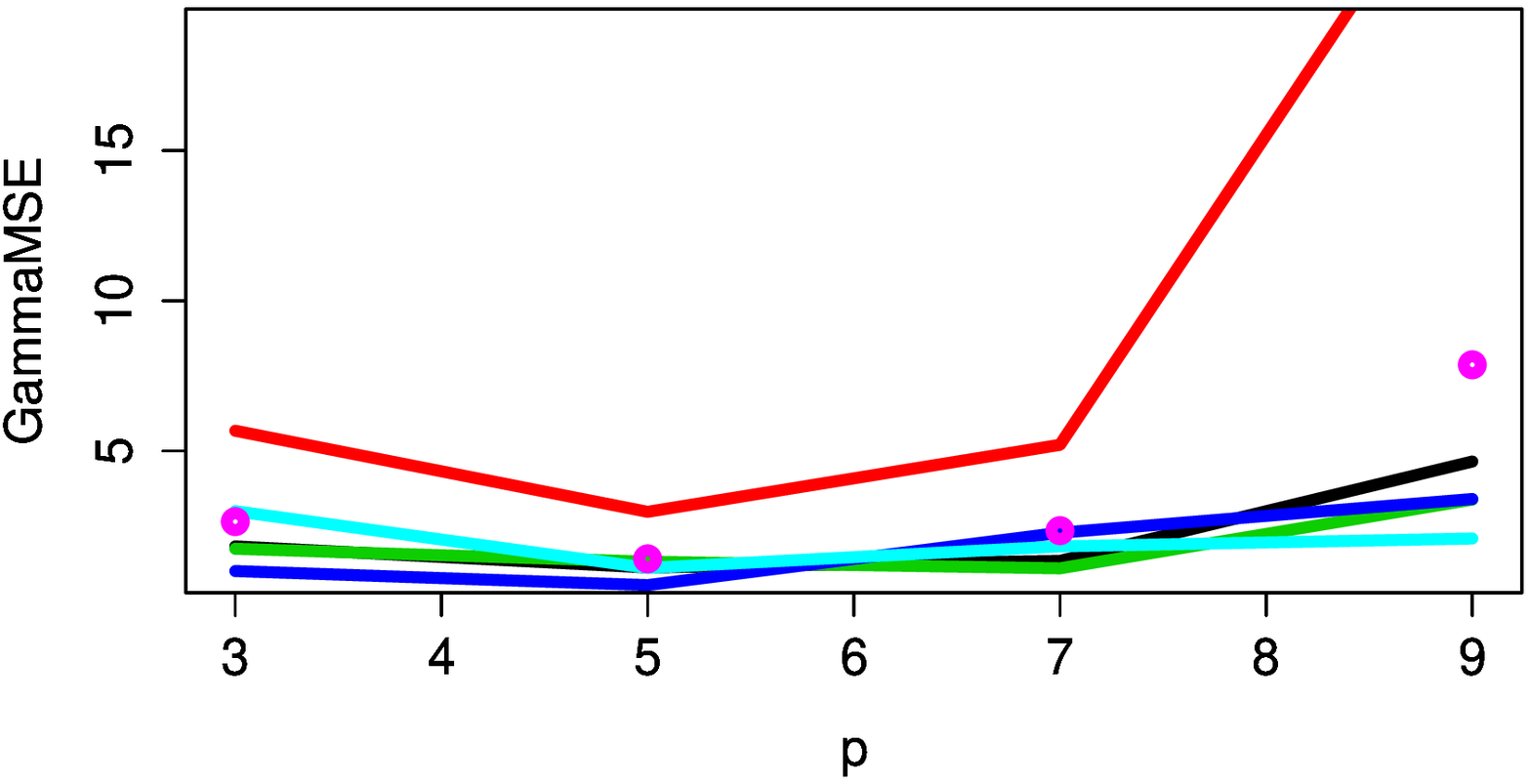}
\tiny{Selection of $p$ for the robust linear model.}
\end{center}
}\\
\parbox{8cm}{
\begin{center}
\includegraphics[width=8cm]{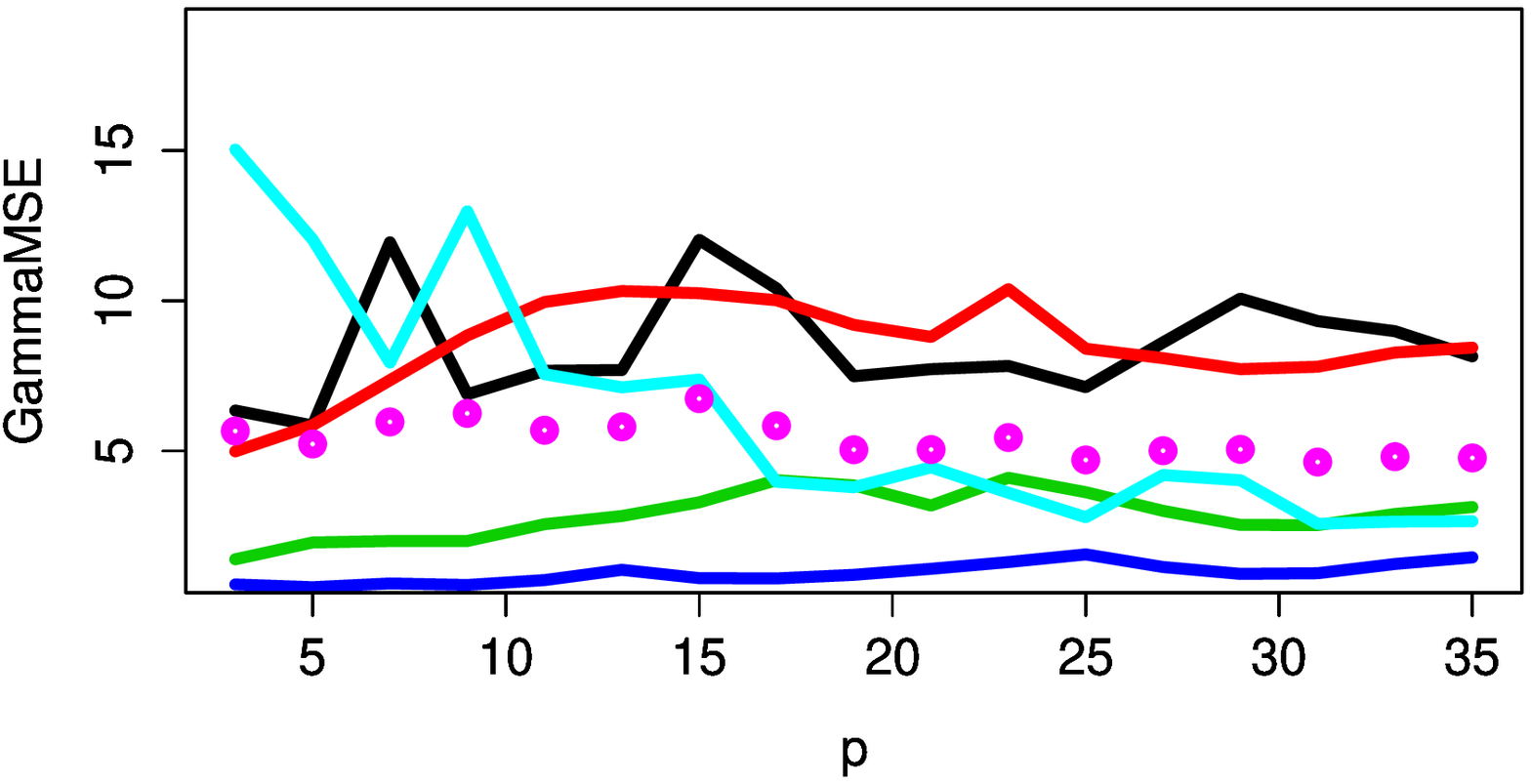}
\tiny{Selection of $p$ for the ridge regression. }
\end{center}
}&
\parbox{8cm}{
\begin{center}
\includegraphics[width=8cm]{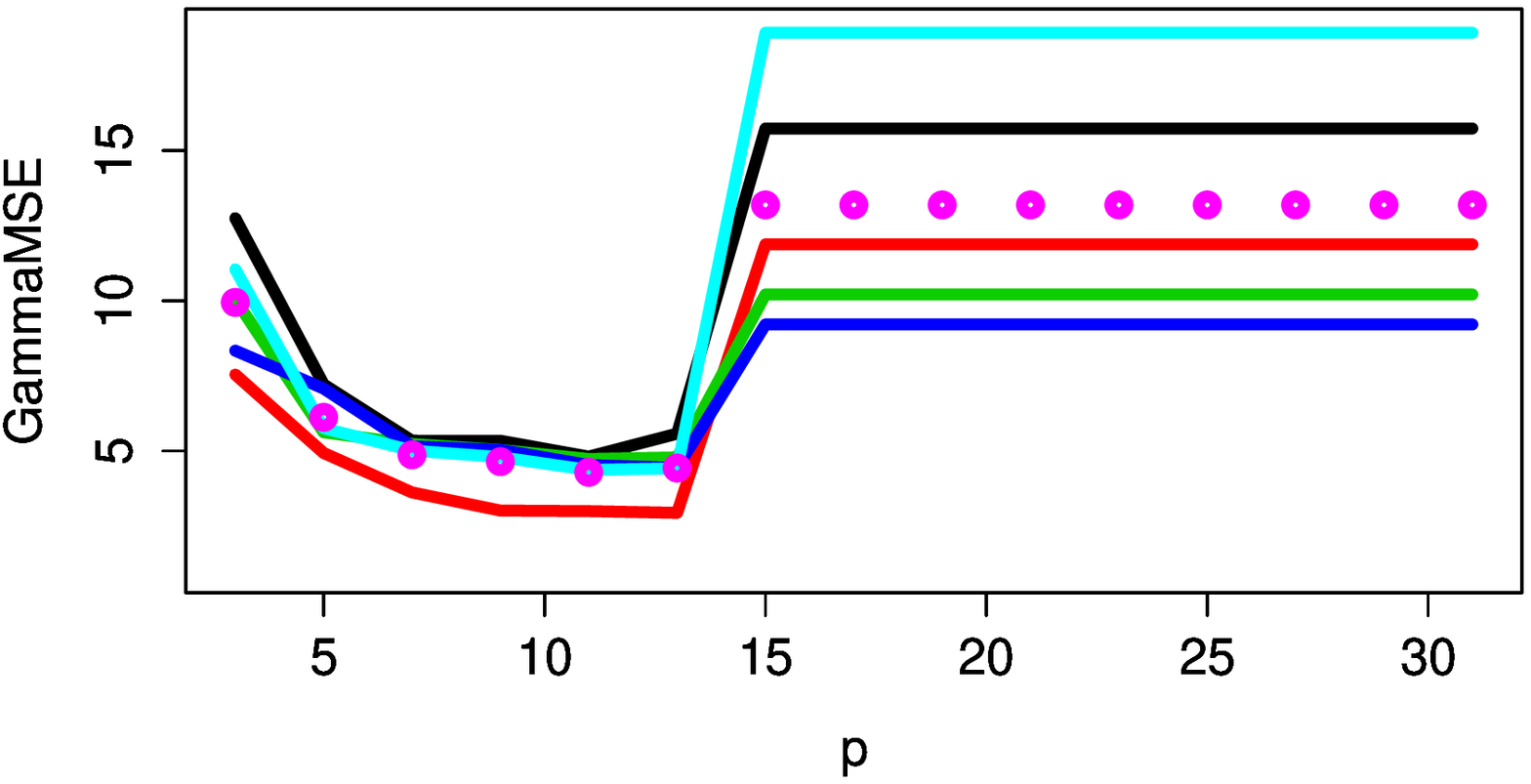}
\tiny{Selection of $p$ for the lasso.}
\end{center}
}

\end{tabular}
\end{center}
\end{figure}

For each procedure, we use a different range for $p$. Notice that $p$ is always odd when we use the Fourier base and must be greater than $4$ when we use the spline decomposition. For the linear model and for robust regression, we recall that $p<13$ due to problems of identification. For ridge regression and the lasso, there are no restrictions and we set $p\leq 31$ for the lasso and $p\leq 35$ for the ridge regression. In figures \ref{figSelp}, we show the evolution of $\Gamma$ME as a function of $p$, the number of Fourier basis functions. Each of the 5 lines corresponds to one validation set and the pink dot corresponds to the mean. For the linear model, the value $p=7$, is clearly optimal. For the robust linear model, the use of 5 basis functions appears marginally preferable. The results for ridge regression do not show a clear trend, however high values for $p$ should be preferred. Finally, the lasso should be used with around 11 Fourier basis functions.

\subsection{Results for prediction}

In Tables \ref{table:regF} and \ref{table:regS}, we detail the numerical accuracy of the prediction procedure, calculated using the validation method explained in \ref{subsec:4.2}. It is clear that all the functional model based methods improve the previous method. For all methods shown, the median number $p$ of basis functions selected (named Nbase in the tables) in the validation step is quite small and is consistent with previous results for functional models \cite{ramsay2}. The robust linear model always outperforms the linear model however the comparison with ridge regression is not so clear. Furthermore, the use of the standardized output variables is better for the linear model but not for the others. We observe that the linear model and the robust linear model fail when the number of basis functions is greater than $13$ because the models have identification problems. Finally, we observe that the results for the decomposition in a Fourier basis are often better than for the spline basis.

\begin{table}
\caption{\label{table:regF}Comparison of methods with Fourier decomposition : prediction.}

\begin{tabular}{cccccccccc}
  \toprule
& &\multicolumn{2}{c}{LM} & \multicolumn{2}{c}{Robust LM} & \multicolumn{2}{c}{Ridge} & \multicolumn{2}{c}{Lasso} \\ 
& Astro & Brut & Norm &  Brut & Norm &  Brut & Norm &  Brut & Norm\\
\midrule

 $RMSE_1$  & 0.171 & 0.088 & 0.061 & 0.062 & 0.070 & 0.080 & 0.065 & 0.123 & 0.109 \\ 
    $RMSE_2$ & 0.143 & 0.089 & 0.088 & 0.060 & 0.081 & 0.075 & 0.074 & 0.109 & 0.084 \\ 
  $\Gamma ME$ & 0.437 & 0.280 & 0.238 & 0.189 & 0.247 & 0.231 & 0.210 & 0.322 & 0.281 \\ 
 Nbase & -- & 7& 7 & 7 & 7 & 7 & 21 & 11 & 11 \\ 
   \bottomrule
\end{tabular}
\end{table}

\begin{table}
\caption{\label{table:regS} Comparison of methods with spline decomposition : prediction. }

\begin{tabular}{lccccccccc}
  \toprule
& &\multicolumn{2}{c}{LM} & \multicolumn{2}{c}{Robust LM} & \multicolumn{2}{c}{Ridge} & \multicolumn{2}{c}{Lasso}\\ 
& Astro & Brut & Norm &  Brut & Norm &  Brut & Norm &  Brut & Norm\\
\midrule
  $RMSE_1$  & 0.171 & 0.119 & 0.116 & 0.116 & 0.120 & 0.079 & 0.060 & 0.122 & 0.103 \\ 
  $RMSE_2$  & 0.143 & 0.102 & 0.102 & 0.096 & 0.112 & 0.075 & 0.067 & 0.103 & 0.094 \\
  $\Gamma ME$ & 0.437 & 0.322 & 0.315 & 0.302 & 0.317 & 0.235 & 0.200 & 0.306 & 0.274 \\  
	Nbase & -- & 5 & 5 & 5 & 5 & 21 & 13 & 9 & 9\\ 
   \bottomrule
\end{tabular}
\end{table}

\subsection{Results for prediction intervals}
In Tables \ref{table:ICF} and \ref{table:ICS}, we present the coverage probabilities of the prediction intervals produced with the procedure explained in \ref{sec:2.3}. We can see that the prediction intervals produced by the linear model and the robust linear model are closer to 0.95 than the one build using the classical method. 
But the prediction intervals are too small when one use the robust regression. This is due to the fact that the robust regression introduce a bias in order to reduce the variance and a bootstrap method using a bias correction should be used is this case. We refer to \cite{Clements01bootstrappingprediction} for such methods.
Finally, it is not so clear that the Fourier basis is better than the spline one; as in the previous case. It depends of the proposed method: Fourier with ridge regression is better than Spline with ridge regression, on the contrary, Spline-LM is better than Fourier-LM.

\begin{table}
\caption{\label{table:ICF} Comparison of methods with Fourier decomposition : coverage probabilities at 95 \% . }

\begin{tabular}{lccccccccc}
  \toprule
& &\multicolumn{2}{c}{LM} & \multicolumn{2}{c}{Robust LM} & \multicolumn{2}{c}{Ridge} & \multicolumn{2}{c}{Lasso}\\ 
& Astro & Brut & Norm &  Brut & Norm &  Brut & Norm & Brut & Norm\\
\midrule
Coverage $Y^{(1)}$ & 0.871 & 0.919 & 0.895 & 0.905 & 0.900 & 0.948 & 0.971 & 0.976 & 0.962 \\ 
Coverage  $Y^{(2)}$ & 0.910 & 0.948 & 0.967 & 0.914 & 0.905 & 0.948 & 0.962 & 0.971 & 0.962 \\  
   \bottomrule
\end{tabular}
\end{table}

\begin{table}
\caption{\label{table:ICS} Comparison of methods with spline decomposition : coverage probabilities at 95 \% . }

\begin{tabular}{lccccccccc}
  \toprule
& &\multicolumn{2}{c}{LM} & \multicolumn{2}{c}{Robust LM} & \multicolumn{2}{c}{Ridge}& \multicolumn{2}{c}{Lasso} \\ 
  & Astro & Brut & Norm &  Brut & Norm &  Brut & Norm &  Brut & Norm\\
	\midrule
Coverage $Y^{(1)}$ & 0.871 & 0.910 & 0.924 & 0.938 & 0.924 & 0.981 & 0.981 & 0.976 & 0.962 \\ 
  Coverage $Y^{(2)}$ & 0.910 & 0.933 & 0.948 & 0.929 & 0.895 & 0.976 & 0.981 & 0.971 & 0.962 \\
   \bottomrule
\end{tabular}
\end{table}

\section{Conclusion}
In this work we have implemented a linear functional data model for the prediction of stellar parameters from their spectra. After the observation (and recording) of a spectrum of a star, astronomers want to extract some stellar parameters, such as stellar radius and effective temperature from it. Normally this is a very time consuming process, where many numerical models must be implemented to generate synthetic spectra for comparing against the observed spectra.
An alternative method is to run a grid of numerical models and then select the  nearest synthetic spectra to the observed one. The approach we propose in this work is to use a linear functional data model to find the nearest grid spectrum to the observed one. This new methodology considerably speeds up the reduction process to obtain stellar parameters and allows us to get the best starting point to perform more numerical calculations in order to get closer synthetic spectra to the observed one.
The Genetic Algorithm method \cite{Mokiem05} takes about one day per spectra, the search in a grid \cite{Lefever07, IACOB} takes about 20 - 30 minutes, while the method proposed here takes just a few seconds.
This article considers a new field of application for functional data analysis. We investigated the accuracy of different functional models for prediction and construction of prediction intervals for astrophysical parameters. We presented a classical cross-validation method in order to select the size of the functional model and to evaluate the performances. We used a bootstrap method to compute the prediction intervals. In the numerical experiments, we saw that these functional-based prediction methods perform far better than the classical method currently used in the astrophysics literature. We noted that the linear regression can be very unstable but this issue is solved both with the robust regression and the ridge regression, improving the prediction performances.
As a future work we will look to create our own grid of synthetic spectra and to use this methodology with a larger number of parameters (temperature, radius, gravity, some elements abundances) to better describe the evolutionary stage of a star.\\

\section{Acknowledgements}
This article was financed by the GEMINI-CONICYT Fund, allocated to the project $N^{\circ} 32120025$. The research of Matthieu Saumard was financed by CONICYT-FONDECYT project  $N^{\circ} 3140602$. MC thanks the support of FONDECYT project 1130173 and Centro de Astrof\'isica de Valpara\'iso.

\bibliographystyle{alpha}
\bibliography{Astronef}

\newcommand{\etalchar}[1]{$^{#1}$}
\begin{thebibliography}{MdKP{\etalchar{+}}05}

\bibitem[BCF11]{cuevas2}
Amparo Ba{\'{\i}}llo, Antonio Cuevas, and Ricardo Fraiman.
\newblock Classification methods for functional data.
\newblock In {\em The {O}xford handbook of functional data analysis}, pages
  259--297. Oxford Univ. Press, Oxford, 2011.

\bibitem[CCF{\etalchar{+}}14]{ciol}
Mattia Ciollaro, Jessi Cisewski, Peter Freeman, Christopher Genovese, Jing Lei,
  Ross O'Connell, and Larry Wasserman.
\newblock Functional regression for quasar spectra.
\newblock {\em http://arxiv.org/abs/1404.3168}, 04 2014.

\bibitem[CFF07]{cuevas}
Antonio Cuevas, Manuel Febrero, and Ricardo Fraiman.
\newblock Robust estimation and classification for functional data via
  projection-based depth notions.
\newblock {\em Comput. Statist.}, 22(3):481--496, 2007.

\bibitem[CFS99]{cardot99}
Herv{{\'e}} Cardot, Fr{{\'e}}d{{\'e}}ric Ferraty, and Pascal Sarda.
\newblock Functional linear model.
\newblock {\em Statist. Probab. Lett.}, 45(1):11--22, 1999.

\bibitem[Cha95]{pikaia}
P.~Charbonneau.
\newblock Genetic algorithms in astronomy and astrophysics.
\newblock {\em Astrophysical Journal Supplement}, 101:309, December 1995.

\bibitem[CK01]{Clements01bootstrappingprediction}
Michael~P. Clements and Jae~H. Kim.
\newblock Bootstrapping prediction intervals for autoregressive models.
\newblock {\em International Journal of Forecasting}, 17:247--267, 2001.

\bibitem[DH12]{del1}
Aurore Delaigle and Peter Hall.
\newblock Achieving near perfect classification for functional data.
\newblock {\em J. R. Stat. Soc. Ser. B. Stat. Methodol.}, 74(2):267--286, 2012.

\bibitem[DH13]{del2}
Aurore Delaigle and Peter Hall.
\newblock Classification using censored functional data.
\newblock {\em J. Amer. Statist. Assoc.}, 108(504):1269--1283, 2013.

\bibitem[DHA09]{del3}
Aurore Delaigle, Peter Hall, and Tatiyana~V. Apanasovich.
\newblock Weighted least squares methods for prediction in the functional data
  linear model.
\newblock {\em Electron. J. Stat.}, 3:865--885, 2009.

\bibitem[FV06]{ferraty}
Fr{\'e}d{\'e}ric Ferraty and Philippe Vieu.
\newblock {\em Nonparametric functional data analysis: theory and practice}.
\newblock Springer, 2006.

\bibitem[HG04]{hamann04}
WR~Hamann and G~Gr{\"a}fener.
\newblock Grids of model spectra for wn stars, ready for use.
\newblock {\em Astronomy \& Astrophysics}, 427(2):697--704, 2004.

\bibitem[HK70]{Hoerl1}
A.~E. Hoerl and R.~W. Kennard.
\newblock Ridge regression: Biased estimation for nonorthogonal problems.
\newblock {\em Technometrics}, 12:55--67, 1970.

\bibitem[Hub64]{huber1964}
Peter~J. Huber.
\newblock Robust estimation of a location parameter.
\newblock {\em Annals of Mathematical Statistics}, 35:73--101, 1964.

\bibitem[HY10]{Selec-hall}
Peter Hall and You-Jun Yang.
\newblock Ordering and selecting components in multivariate or functional data
  linear prediction.
\newblock {\em J. R. Stat. Soc. Ser. B Stat. Methodol.}, 72(1):93--110, 2010.

\bibitem[KP00]{KP2000}
R.-P. Kudritzki and J.~Puls.
\newblock Winds from hot stars.
\newblock {\em Annual Review of Astronomy and Astrophysics}, 38:613--666, 2000.

\bibitem[LPA07]{Lefever07}
K.~Lefever, J.~Puls, and C.~Aerts.
\newblock Statistical properties of a sample of periodically variable b-type
  supergiants. evidence for opacity-driven gravity-mode oscillations.
\newblock {\em Astronomy \& Astrophysics}, 463:1093--1109, March 2007.

\bibitem[LWC13]{selec-LWC}
Yehua Li, Naisyin Wang, and Raymond~J. Carroll.
\newblock Selecting the number of principal components in functional data.
\newblock {\em J. Amer. Statist. Assoc.}, 108(504):1284--1294, 2013.

\bibitem[MAK08]{matsui08}
Hidetoshi Matsui, Yuko Araki, and Sadanori Konishi.
\newblock Multivariate regression modeling for functional data.
\newblock {\em J. Data Sci}, 6(3):313--331, 2008.

\bibitem[MdKP{\etalchar{+}}05]{Mokiem05}
M.~R. Mokiem, A.~de~Koter, J.~Puls, A.~Herrero, F.~Najarro, and M.~R.
  Villamariz.
\newblock Spectral analysis of early-type stars using a genetic algorithm based
  fitting method.
\newblock {\em Astronomy \& Astrophysics}, 441:711--733, October 2005.

\bibitem[Rob07]{Rob07}
K.~Robinson.
\newblock {\em Spectroscopy: The Key to the Stars}.
\newblock Springer, 2007.

\bibitem[RS02]{ramsay2}
James~O Ramsay and Bernard~W Silverman.
\newblock {\em Applied functional data analysis: methods and case studies},
  volume~77.
\newblock Springer, 2002.

\bibitem[RS05]{ramsay}
JO~Ramsay and BW~Silverman.
\newblock {\em Functional data analysis.}
\newblock Springer, New York, 2005.

\bibitem[SDH14]{IACOB}
S.~Sim{\'o}n-D{\'{\i}}az and A.~Herrero.
\newblock The iacob project. i. rotational velocities in northern galactic o-
  and early b-type stars revisited. the impact of other sources of
  line-broadening.
\newblock {\em Astronomy \& Astrophysics}, 562:A135, February 2014.

\bibitem[SH12]{hsing}
Hyejin Shin and Tailen Hsing.
\newblock Linear prediction in functional data analysis.
\newblock {\em Stochastic Process. Appl.}, 122(11):3680--3700, 2012.

\bibitem[Sti85]{Stine85}
R.~A. Stine.
\newblock Bootstrap prediction intervals for regression.
\newblock {\em Journal of the American Statistical Association}, 80:1026--1031,
  1985.

\bibitem[Tib96]{tibshirani96regression}
R.~Tibshirani.
\newblock Regression shrinkage and selection via the lasso.
\newblock {\em Journal of the Royal Statistical Society (Series B)},
  58:267--288, 1996.

\end{thebibliography}

\end{document}